\documentclass[conference]{IEEEtran}
\IEEEoverridecommandlockouts

\usepackage[T1]{fontenc}
\usepackage{graphicx}
\usepackage{bm}

\usepackage{amsmath,amsfonts,bm}

\def\eqref#1{equation~\ref{#1}}

\def\1{\bm{1}}

\def\vh{{\bm{h}}}

\def\vo{{\bm{o}}}

\def\vq{{\bm{q}}}
\def\vr{{\bm{r}}}

\def\vx{{\bm{x}}}
\def\vy{{\bm{y}}}
\def\vz{{\bm{z}}}

\def\mF{{\bm{F}}}

\def\mH{{\bm{H}}}
\def\mI{{\bm{I}}}

\def\mK{{\bm{K}}}

\def\mQ{{\bm{Q}}}
\def\mR{{\bm{R}}}

\def\mSigma{{\bm{\Sigma}}}

\DeclareMathAlphabet{\mathsfit}{\encodingdefault}{\sfdefault}{m}{sl}
\SetMathAlphabet{\mathsfit}{bold}{\encodingdefault}{\sfdefault}{bx}{n}

\newcommand{\C}{\mathbb{C}}

\usepackage{todonotes}

\def\defeq{:=}

\usepackage{color, soul}
\usepackage[printonlyused]{acronym}
\acrodef{AR}{auto-regressive}
\acrodef{ARMA}{auto-regressive moving average}
\acrodef{CP}{cyclic prefix}
\acrodef{CDL}{clustered delay line}
\acrodef{KF}{Kalman Filter}
\acrodef{MMSE}{Minimum Mean Square Error}
\acrodef{MNSE}{Mean Normalized Square Error}
\acrodef{NSE}{Normalized Square Error}
\acrodef{NN}{Neural Network}
\acrodef{HKF}{Hypernetwork Kalman Filter}
\acrodef{BKF}{Binned Kalman Filter}
\acrodef{GKF}{genie \ac{KF}}
\acrodef{UE}{User Equipment}
\acrodef{ISI}{Intersymbol Interference}

\begin{document}
\title{
Neural Augmentation of Kalman Filter  with Hypernetwork for Channel Tracking
}
\author{
    \IEEEauthorblockN{Kumar Pratik\IEEEauthorrefmark{1}, Rana Ali Amjad\IEEEauthorrefmark{1}, Arash Behboodi\IEEEauthorrefmark{1}, Joseph B. Soriaga\IEEEauthorrefmark{2}, Max Welling\IEEEauthorrefmark{1}}
    \IEEEauthorblockA{\IEEEauthorrefmark{1}Qualcomm Technologies Netherlands B.V.}
    \IEEEauthorblockA{\IEEEauthorrefmark{2}Qualcomm Technologies, Inc.}
    \IEEEauthorblockA{Qualcomm AI Research}
\thanks{Rana Ali Amjad completed the research that is the basis of this paper during employment at Qualcomm Technologies Netherlands B.V.}
\thanks{Qualcomm AI Research is an initiative of Qualcomm Technologies, Inc.}
}

\maketitle

\begin{abstract}
We propose \ac{HKF} for tracking applications with multiple different dynamics. The \ac{HKF} combines generalization power of Kalman filters with expressive power of neural networks. 
Instead of keeping a bank of Kalman filters and choosing one based on approximating the actual dynamics, \ac{HKF} adapts itself to each dynamics based on the observed sequence. 
Through extensive experiments on CDL-B channel model, we show that the \ac{HKF} can be used
for tracking the channel over a wide range of Doppler values, matching Kalman filter performance with genie Doppler information. At high Doppler values, it achieves around 2dB gain over genie Kalman filter.  The \ac{HKF} generalizes well to unseen Doppler, SNR values and pilot patterns unlike LSTM, which suffers from severe performance degradation.
\end{abstract}

\section{Introduction}
Channel tracking in wireless communication leverages the knowledge about the dynamics of the time varying channels to improve channel estimation quality. The channel dynamics is determined by the Doppler frequency. When the Doppler frequency is known, \ac{KF} is widely used for tracking \cite{barbieri_arma_2009,kashyap_performance_2017,el_husseini_optimization_2017}. An \ac{AR} model is assumed for the transition dynamics, and the parameters are chosen either based on a Doppler dependent model, e.g., Jakes model or by fitting the parameters to the data. \ac{KF} is MMSE optimal when the transition dynamics, observation model, and noise statistics follow a linear Gaussian assumption. It can elegantly adapt to missing observations and is robust to noise variation. If the underlying dynamics changes, Kalman parameters need to be updated according to the new dynamics. 
In practice, when the Doppler value can change over a wide range, the space of possible dynamics are roughly quantized into different bins, and a finite number of \ac{KF}s, one per Doppler bin. are stored . 
This process is prone to error propagation. Any mistake in approximating the Doppler value can lead to a wrong choice of Kalman and incur considerable loss. 

During recent years, recurrent neural networks (RNNs) have appeared as a promising solution for tracking application. 
With high expressive and approximation power, these models can be trained on a large dataset of all possible scenarios with the hope that a single neural network model can smoothly interpolate between different operation regimes and replace the bank of Kalman filters. However, in this paper, we will show that a naive application of these methods to channel tracking suffers from various issues. Their performance degrades significantly on unseen cases with  deviations from the training scenario. 

Inspired by \cite{satorras_combining_2019}, we adopt a hybrid approach. Instead of replacing the \ac{KF} by \acp{NN}, we keep the underlying graphical model and only update the parameters of Kalman using a hypernetwork. The \acl{HKF} uses only Kalman equations for prediction and updates its parameters continuously based on the dynamics of the channel, which is learned online from the past. In this way, the model inherits the benefits of Kalman, for instance handling of missing observations or varying SNR. 

Our contributions are as follows. We propose \ac{HKF} for tracking channels with unknown and varying dynamics. At each time step, a neural network updates the parameters of the \ac{KF} based on the latent representation of past sequence. The prediction and tracking are done using Kalman equations. We evaluate this model for tracking channel taps in an OFDM transmission. We have used the \ac{CDL} channel model from 3GPP standard \cite{3gpp_38900_2016}. We show that a single LSTM suffers significantly from model mismatch. In contrast, the proposed \ac{HKF} consistently outperforms LSTM and provides gain over Kalman across a wide range of Doppler.


\subsection{Related Works}

Channel tracking is important for continuous transmission in time varying channels. Kalman filtering is the standard tool, and there are many papers around this problem (see for instance \cite{barbieri_arma_2009,kashyap_performance_2017,el_husseini_optimization_2017}). In these works, the underlying dynamics of the channel is known, therefore the Kalman parameters can be matched to the dynamics. However, we assume a multi-Doppler scenario where the Doppler is not known a priori. The authors in \cite{satorras_combining_2019}
proposed a non-causal hybrid model where the Kalman updates are modeled as message passing algorithm learned using \acp{NN}. In contrast, in our work, Kalman updates are not modeled by an \ac{NN}, and the model is causal. 

There are many works related to tracking and forecasting which leverage \ac{NN}s to learn hidden state space model~\cite{karl_deep_2016, krishnan2017structured, de_bezenac_normalizing_2020, coskun2017long}. While most of the previous works focus on making \ac{KF} more complex, e.g., non-linear Kalman transitions, the closest work to our paper is \cite{rangapuram_deep_2018} where the authors similarly use an LSTM to update Kalman parameters. However, they don't have missing observations in the training sequence. During inference they deal with missing observation by assuming that there are always some covariates available that are correlated with the observation. This is not the case in wireless communication. Between pilot transmissions, there is no covariate observation available. 



\section{Problem Setup}
\label{sec:prob_setup}

Consider an OFDM system with $N$ sub-carriers. The communication spans over $T$ consecutive OFDM symbols. At OFDM symbol $t$, the source signal $\vx_t\in\C^N$  is modulated over $N$ sub-carriers using IFFT operation and is transmitted after \ac{CP} addition. We assume that \ac{CP} is long enough to remove \ac{ISI}. However, the channel is changing in time with a Doppler frequency $f_d$. We assume a multi Doppler scenario where  the Doppler frequency can differ from one scenario to another. The channel is estimated using known pilot OFDM symbols at some intervals. The pilots are transmitted once every $T_{p}$ OFDM symbols. We assume that known QPSK symbols are modulated over all sub-carriers in each pilot OFDM symbols. The channel at time $t$ is denoted by $\vh_t\in\C^N$. The estimated channel solely based on the pilot at time $t$ is denoted by $\vo_t$ given as
\begin{equation}
        {\vo}_t=\vh_t+\vr_t,
\label{eq:observation}
\end{equation}
where $\vr_t$ is the additive noise with the covariance matrix $\mR_t$. 

The goal is to track the channel between pilot transmissions and also use past information to improve the estimated channel from pilots. The final estimated channel at time $t$ is denoted by $\hat{\vh}_{t}$. The error for each instance is measured using \ac{NSE} defined as:
\begin{equation}
\text{NSE}(t)=\frac{\|\hat{\vh}_t-\vh_t\|_2^2}{\|\vh_t\|_2^2}
\label{eq:MNSE}
\end{equation}
The final reported error is the average \ac{NSE}, namely \ac{MNSE}. Two remarks are in order. First, to reduce the dimension, we track the channel in time domain in our experiments. Second, although we have considered SISO channels, the method can be easily extended to MIMO channels.

\subsection{Kalman based Channel Tracking}
\label{subsec:kalman}
We start by presenting Kalman filter based solution to this problem. For details of Kalman equations, we refer to classical textbooks like \cite{kailath_lectures_2014}. We use \ac{AR} models for the transition dynamics of $\vh_t$. Particularly, we use AR(2) Kalman equations given by:
\begin{align}
 \begin{pmatrix}
    {\vh}_t\\
    {\vh}_{t-1}
    \end{pmatrix} & =\mF_t  \begin{pmatrix}
    {\vh}_{t-1}\\
    {\vh}_{t-2}
    \end{pmatrix} +\begin{pmatrix}
       \vq_t\\
       \mathbf{0}
    \end{pmatrix},
\label{eq:process_equation}\\
 \begin{pmatrix}
    {\vo}_t\\
    {\vo}_{t-1}
    \end{pmatrix} & = \mH_t  \begin{pmatrix}
    {\vh}_{t}\\
    {\vh}_{t-1}
    \end{pmatrix} +\begin{pmatrix}
       \vr_t\\
       \vr_{t-1}
    \end{pmatrix}.
\label{eq:obs_equation}
\end{align}
The matrix $\mF_t$ models the transition dynamics, and $\mH_t$ models the observation matrix. For AR(2)-Kalman based channel tracking, we have: 
\[
\mF_t\defeq\begin{pmatrix}
    {\mF}_1^t&\mF_2^t\\
    {\mI}& \mathbf{0}
    \end{pmatrix}, \mH_t \defeq \mI.
\]
The vector $\vq_t$ is the process noise with the covariance matrix $\mQ_t$. By $\tilde{\mQ}_t$, we denote the covariance matrix of the total noise vector in \eqref{eq:process_equation}.
    The covariance matrix of total observation noise $\begin{pmatrix}
       \vr_t&
       \vr_{t-1}
    \end{pmatrix}^T$ is denoted by $\tilde{\mR}_t$. Note that by assuming $\mF^t_2=0$, we get the AR(1) model.

Kalman updates can be computed using \eqref{eq:process_equation} and \eqref{eq:obs_equation}. We have to consider two cases for channel estimation, first for information symbols where no observation is given, and next, for  pilot symbols where the observation $\vo_t$ is present. At time $t$, we have access to previous estimates $\hat{\vh}_{t-1}$ and $\hat{\vh}_{t-2|t-1}$\footnote{Throughout the text, $\vh_{s|t}$ means the estimate of $\vh_s$ at time $t$.} and the covariance matrix of the estimate $    \begin{pmatrix}
    \hat{\vh}_{t-1}\\
    \hat{\vh}_{t-2|t-1}
    \end{pmatrix}$ denoted by $\mSigma_{t-1|t-1}$. When no observation, i.e., no pilot is available at time $t$, the estimated channel is equal to:
\begin{equation}
    \begin{pmatrix}
    \hat{\vh}_{t}\\
    \hat{\vh}_{t-1|t}
    \end{pmatrix} ={\mF}_t\begin{pmatrix}
    \hat{\vh}_{t-1}\\
    \hat{\vh}_{t-2|t-1}
    \end{pmatrix} .
\label{eq:KF_missing_obs}
\end{equation}
Note that with AR(2)-model, $\hat{\vh}_{t-1|t}$ is simply equal to $\hat{\vh}_{t-1}$.
When the observation $\vo_t$ is available, the estimated channel $\hat{\vh_t}$ is given recursively by Kalman updates:
\begin{equation}
    \begin{pmatrix}
    \hat{\vh}_{t}\\
    \hat{\vh}_{t-1|t}
    \end{pmatrix}={\mF}_t\begin{pmatrix}
    \hat{\vh}_{t-1}\\
    \hat{\vh}_{t-2|t-1}
    \end{pmatrix}
    +\mK_t\vy_t,
    \label{eq:KF_with_obs}
\end{equation}
where $\mK_t$ and $\vy_t$ are respectively the Kalman gain and the Kalman innovation. The Kalman innovation is given by:
\begin{equation}
    \vy_{t}= \begin{pmatrix}
    {\vo}_t\\
    {\vo}_{t-1}
    \end{pmatrix}-{\mH}_t{\mF}_t\begin{pmatrix}
    \hat{\vh}_{t-1}\\
    \hat{\vh}_{t-2|t-1}
    \end{pmatrix}.
\end{equation}
The Kalman gain is recursively computed from the estimate covariance matrix of last time step denoted by $\mSigma_{t-1|t-1}$.
\begin{equation}
    \mK_t=\mSigma_{t|t-1}\mH_t^H(\mH_t\mSigma_{t|t-1}\mH_t^H+\tilde{\mR}_t)^{-1},
\end{equation}
where
$
\mSigma_{t|t-1}=\mF_t\mSigma_{t-1|t-1}\mF_t^H+\tilde{\mQ}_t.
$
The estimate covariance matrix at time $t$, $\mSigma_{t|t}$ is given by $\mSigma_{t|t-1}$ if there is no new observation. Otherwise, $\mSigma_{t|t}$ is given by $(\mI- \mK_t\mH_t)\mSigma_{t|t-1}$. For \ac{AR}(2) model, we use the assumption that $\mH_t=\mI$, which simplifies the above equations. These equations can be recursively computed with known $(\mF_1^t,\mF_2^t,\mQ_t,\mR_t)$ for all $t$. In this paper, since the ground truth channel is known in training time, Kalman parameters can be obtained by a simple linear regression.  
For multi Doppler case, \ac{KF} parameters should be chosen according to the actual Doppler. This requires keeping a bank of \ac{KF}s at hand with an additional Doppler estimation unit. In this paper, we assume that the Doppler range is divided into finite bins with one \ac{KF} per bin. We assume that Doppler frequency is known for choosing the correct bin. We refer to this approach as \ac{BKF}. For each bin, the parameters of the \ac{KF} is computed over the pre-selected Doppler values in the bin. For example see Table \ref{tab:binning_strategy}.

\subsection{RNN based Channel Tracking}
\label{subsec:rnn}
To avoid the overhead of explicit Doppler estimation and maintenance of a bank of \ac{KF}s, we can adopt a data driven approach. As an alternative to \ac{KF}, we can use an RNN for prediction, which is trained over the full range of Doppler values. Therefore, a single model can replace bank of \ac{KF}s without any need for explicit Doppler estimation. To account for missing observations, we make the RNN predict, at every time step, the current channel estimate $\hat{\vh}_t$ and the next observation $\hat{\vo}_{t+1}$. In case of missing observation, the RNN can take this synthetic estimated observation $\hat{\vo}_{t+1}$ as input. We use the real observation $ \vo_{t+1}$ whenever it is present. We define $\tilde{\vo}_t$ as follows:
\begin{equation}
\tilde{\vo}_t=\begin{cases}
\vo_t & \text{observation is present} \\
\hat{\vo_t} & \text{missing observation}
\end{cases}
\label{eq:synth_obs}    
\end{equation}

The recurrent iterations of the RNN is given by:
    \begin{equation}
        \vz_t = \text{RNN}(\vz_{t-1}, \hat{\vh}_{t-1}, \tilde{\vo}_t)
    \end{equation}
    \begin{equation}
        \hat{\vh}_t, \hat{\vo}_{t+1} = \text{MLP}(\vz_t)
    \end{equation}
    where, $\vz_t$ represents the state variable of the RNN, $\hat{\vh}_t$ represents the channel estimate at time $t$, and $\hat{\vo}_{t+1}$ represents the synthetic observation for time $t+1$. The loss function used to train the RNN is given by:
    \begin{equation}
        \mathcal{L}_{RNN}(\phi) = \sum_{m=1}^{M}\sum_{t=1}^{T}MSE(\hat{\vh}_t^{m}(\phi), \vh_t^{m}) + MSE(\hat{\vo}_t^{m}(\phi), \vo_{t}^{m}).
    \end{equation}
    Here, $m$ denotes different the training sample index, and $t$ denotes the time entries in the sequence. $T$ represents the sequence length. $MSE$ is the mean squared error, and $\phi$ represents the trainable RNN and MLP parameters.
    
    RNNs have properties that are complementary to the \ac{KF}. Unlike \ac{KF}, RNNs do not assume any linear or Gaussian constraints on the transition and observation dynamics. Also, the RNNs do not require the evolution dynamics to be stationary as it can learn to extract the time varying dynamics directly from the training data. At the same time, RNNs have their own share of limitations. Like most of the deep learning methods, RNNs are very sensitive to the training data and do not generalize well to the settings other than what it is trained for, which makes them brittle for real world deployment. 
    The RNNs are found to be notoriously difficult to train end-to-end due to the infamous exploding and vanishing gradient problem. On the opposite end of the spectrum, \ac{KF} models time sequences in a very interpretable manner by explicitly depicting the transition model, observation model, and noise model parameters. This elegant structure of \ac{KF} renders it with properties inexistent in solely \ac{NN} based solutions, like efficient handling of missing entries in the time sequence and robustness to out of distribution data.
    As we will see in the section~\ref{sec:experim}, the RNN based tracking scheme fails to achieve performance competitive to the \ac{BKF} and also faces difficulties in generalization to unseen scenarios.  In the next section, we discuss the technical details of the proposed \ac{HKF}, which can overcome this problem.

\section{Hypernetwork Kalman Filter}
\label{sec:hyperKF}

\begin{figure}
            \centering
            \includegraphics[width=0.51\textwidth]{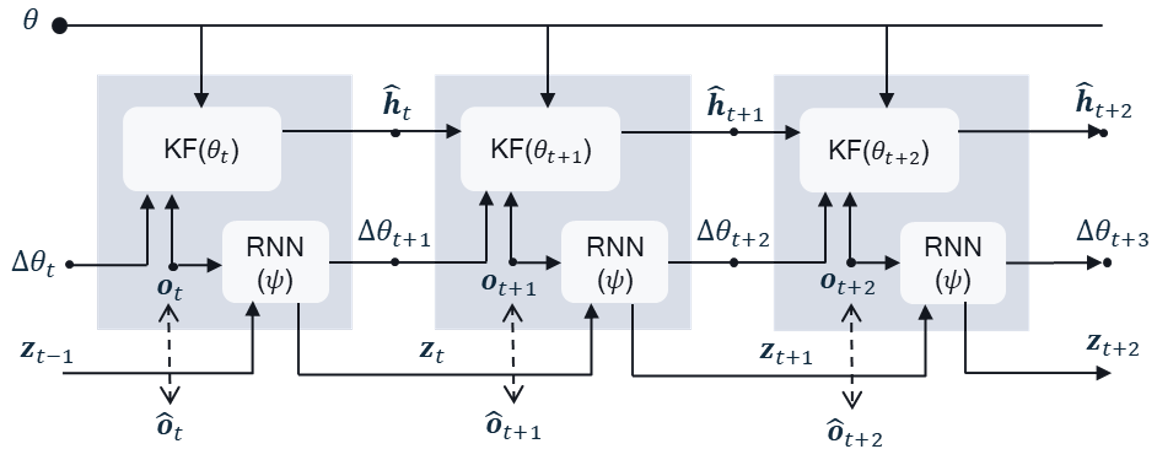}
            \caption{\textit{Illustration of Hypernetwork Kalman Filter.} Each shaded block represents one inference iteration of the \ac{HKF}, where $\theta$ denotes the base set of \ac{KF} parameters around which the RNN models the residual $\Delta \theta_t$, $\theta_t$ is the final \ac{KF} parameters at step $t$, $\vz_t$ is the RNN state variable, $\vo_t$ and $\hat{\vo}_t$ are the real and synthetic observation respectively, and $\psi$ represents the trainable parameters of the \ac{HKF}. The RNN extracts the time varying dynamics of the underlying process and informs the \ac{KF}, at every step, with the optimal set of  \ac{KF} parameters.}
            \label{fig:hypernetwork_kf}
        \end{figure}

   The proposed \ac{HKF} complements the flexibility of the RNNs in learning to extract the dynamics from the data with the robustness and interpretability of the \ac{KF}. We extend the class of evolutionary processes that a \ac{KF} can model by augmenting it with an RNN. The \ac{HKF} retains the interpretability of \ac{KF} and at the same time circumvents the limitations posed by a standalone \ac{KF} by incorporating an RNN whose parameters are learned from the training data. 
    The \ac{HKF} consists of a Kalman filter accompanied by an RNN to augment its capabilities. The prediction is still done by Kalman, thereby enjoying robustness and generalization of Kalman. However, Kalman parameters are updated at each time using an RNN based on the process history. A detailed schematic of the \ac{HKF} is depicted in Fig.~\ref{fig:hypernetwork_kf}.  Below we describe the details of both the constituent units of the HKF: the Kalman filter and the Hypernetwork RNN.
    
    \subsection{Kalman Filter}
        At the core of \ac{HKF} lies a classical Kalman filter. The details of Kalman is given in the previous section.
        As we have seen, a Kalman filter is parameterized completely by the parameter set $\theta_t\defeq(\mF_1^t, \mF_2^t, \mH_t, \mQ_t, \mR_t)$. In classical time-stationary \ac{KF}, $\theta_t$ is assumed to be the same at each time step which immensely limits the class of evolutionary dynamics it can model. The traditional binning based channel tracking method circumvents this limitation by coarsely binning the bigger set of possible dynamics into subsets and maintaining a different set of \ac{KF} parameters per bin.  In the proposed \ac{HKF}, the parameters $\theta_t$ at each time $t$ are updated by the hypernetwork RNN. 
        
    \subsection{Hypernetwork RNN}
        As illustrated in Fig.~\ref{fig:hypernetwork_kf}, at every time step $t$ the RNN updates the \ac{KF} parameters ($\theta_{t+1}$) for the next time step $t+1$. The RNN models the  \ac{KF} parameters in terms of residual around the mean set of parameters $\theta$. In other words, the Kalman base parameters are fixed to $\theta$, and the RNN provides corrections: 
        \begin{equation}
            \vz_t = \text{RNN}(\vz_{t-1}, \tilde{\vo}_t), \text{    } \Delta \theta_{t+1} = \text{MLP}(\vz_t)
        \end{equation}
        \begin{equation}
            \theta_{t+1} = \theta + \Delta \theta_{t+1}
        \end{equation}
        The hidden state of RNN is projected by a single layer MLP (zero hidden layer) to the required dimension which is then recasted into $\ac{KF}$ parameters domain. Note that $\tilde{o}_t$ is defined similar to \eqref{eq:synth_obs}. In case of missing observation ($\vo_t$), the RNN uses the \ac{KF} estimate at time $t$, i.e., $\hat{\vh}_{t}$ and forward it through the observation process ($\mH_t, \mR_t$). This means that the synthetic observation $\hat{\vo}_t$ is modeled as a Gaussian random vector with mean value $\mH_t\hat{\vh}_t$ and the covariance $\mR_t$. The synthetic observation is then fed into the RNN. To be able to backpropagate through this sampling process, we use the reparameterization trick~\cite{kingma2013auto}:
        \begin{equation}
            \hat{\vo}_t = \mH_t \hat{\vh}_t + \mR_t^{1/2} \pmb{\varepsilon}, \text{  } \pmb{\varepsilon} \sim \mathcal{N}(0,\mathbb{I}_n) 
        \end{equation}
        In the channel tracking problem, we assume that $\mH_t$ is set to identity and $\mR_t$ is a diagonal matrix with diagonal elements given by the vector $\mR_{diag}$. Then, the sampling operation can be simplified to $\mR_{diag}^{1/2} \odot \pmb{\varepsilon}$.
        
    \subsection{Hypernetwork Kalman filter}
        The \ac{HKF}, at every step, has access to the base set of \ac{KF} parameters. The stationary base parameters are given by  $\theta = (\mF_1, \mF_2, \mH, \mQ, \mR)$. The hypernetwork RNN models the correction term for them. In our experiments for channel tracking, we assume a perfect knowledge of the measurement process, i.e., the measurement matrix $\mH$ is identity, and the SNR is perfectly known/estimated. Therefore, the observation noise covariance matrix $\mR$ is diagonal with entries that are determined by the genie SNR value. The \ac{KF} parameters which varies across different Doppler scenarios are $\mF_1^t, \mF_2^t,$ and $\mQ_t$. Hence, the RNN only models residuals in these parameters, i.e., $\Delta \theta_{t+1} = \{\Delta \mF_1^{t+1}, \Delta \mF_2^{t+1}, \Delta \mQ_{t+1}\}$. In our experiments, we set $\mF_1$ to identity matrix, $\mF_2$ to zero matrix, and $\mQ$ to \ac{KF} process covariance matrix averaged across the entire training dataset. The loss function used to train the \ac{HKF} is given by:
        \begin{equation}
            \mathcal{L}_{HKF}(\psi) = \sum_{m=1}^{M}\sum_{t=1}^{T}MSE(\hat{\vh}_t^{m}(\psi), \vh_t^{m})
        \end{equation}
        Here, $\psi$ represents the trainable parameters of the \ac{HKF} (hypernetwork RNN and the associated MLP).
    
\section{Experiments}
\label{sec:experim}
    For channel tracking, we have generated a dataset consisting of $15$ Doppler values and then binned them into $5$ mutually exclusive and collectively exhaustive bins. For binning, we use the genie Doppler information to use the correct bin for each channel sequence. We use MATLAB for dataset generation with CDL-B channel profile. We have used $4096$ tones for OFDM transmission at $4$ GHz carrier frequency, delay spread of $100$ ns, and sub-carrier spacing of $30$ KHz. Table~\ref{tab:binning_strategy} depicts the binning strategy used for our experiments, i.e., the Doppler values per bin followed by the corresponding \ac{UE} relative velocity. Our dataset contains $800$ training channel instances, and $200$ test channel instances per Doppler frequency. Each channel instance is $1500$ OFDM symbols long. Unless specified otherwise, we use an SNR of $10$ dB and a pilot ratio of $1:6$, i.e.,  a noisy observation at every $6^{th}$ time step. The evaluation metric in our experiments is MNSE as in \eqref{eq:MNSE} measured in dB unit. Although any RNN can be used, we have used LSTM for the rest. To ensure a fair comparison across different learning based methods, we keep the size of the network hidden state ($z$) at $2N$ for both the RNN and the hypernetwork RNN.
    
    In terms of computational complexity, the \ac{HKF} has the highest complexity followed by the RNN, further followed by the variants of \ac{KF}. It should be noted that non-learning based methods, i.e., \ac{HKF}, and \ac{BKF}, have an extra computational overhead due to the Doppler estimation unit for Doppler shift estimation but the \ac{NN} based methods doesn't need explicit Doppler information as they learn to extract the channel dynamics implicitly from the pilots.
    
    \subsection{Evaluation of different channel tracking schemes}
    In the first experiment, we train one \ac{HKF} on the entire dataset consisting of all the $15$ Doppler values. To compare \ac{HKF} against a standalone \ac{NN} based baseline, we also train an LSTM on the similar settings. We compare both of these learned methods with \ac{GKF} and \ac{BKF}. The parameters of \ac{GKF} is matched exactly to the Doppler value and obtained using the data generated from the same Doppler. Note that this approach would be prohibitively complex in practice, as we require to keep in infinite bank of Kalman filters, one for each Doppler. On the other hand, as mentioned before, the \ac{BKF} is an intermediate solution, which only keeps a finite bank of Kalman filters, each one corresponding to a Doppler range. We have chosen five bins presented in Table \ref{tab:binning_strategy}. Table~\ref{tab:eval_table} charts the evaluation results on the test data for all the $15$ Doppler values. Our \ac{HKF} performs consistently better than the standalone LSTM. A single \ac{HKF}, without any genie Doppler information, outperforms the \ac{BKF} baseline which requires genie Doppler information and $5$ separate \ac{KF}s. At higher Doppler frequencies, the \ac{HKF} even outperforms the \ac{GKF}, which uses a separate \ac{KF} per Doppler. 
    

    \begin{table}[t]
    \vspace{4mm}
        \centering
            \begin{tabular}{|c|c|c|} 
            \hline
            \multicolumn{3}{|c|}{Bin setup} \\
            \hline
            Bin index & Doppler values (Hz) & corresponding velocity (kmph)\\ [0.5ex] 
            \hline
            Bin 0 & 0, 30, 60  & 0.0, 8.0, 16.0\\ 
            \hline
            Bin 1 & 70, 100, 130 & 18.0, 27.0, 35.0\\
            \hline
            Bin 2 & 150, 210, 270 & 40.5, 56.6, 72.8\\
            \hline
            Bin 3 & 300, 400, 500 & 81.0, 108.0, 135.0\\
            \hline
            Bin 4 & 800, 1300, 1850 & 215.8, 350.7, 499.0\\
            \hline
        \end{tabular} \\[0.5ex]
        \caption{\textnormal{Bin based discretization of Doppler spectrum}}
        \label{tab:binning_strategy}
        \vspace{-4mm}
    \end{table}


    \begin{table}[ht]
        \centering
            \begin{tabular}{||c|c|c|c|c|c||} 
            \hline
            \multicolumn{6}{|c|}{MNSE (in dB)} \\
            \hline
            Doppler & GKF & BKF & LSTM & HKF & HKF-2\\ [0.5ex] 
            \hline\hline
            0 Hz & \textbf{-48.89} & -31.78 & -18.05 & -29.99 & -31.86\\ 
            \hline
            30 Hz & \textbf{-32.60} & \textbf{-32.60} & -22.16 & -30.59 & -30.62\\
            \hline
            60 Hz & \textbf{-31.40} & -28.47 & -26.77 & -30.76 & -30.92\\
            \hline
            70 Hz & -30.64 & -28.84 & -26.63 & -30.75 & \textbf{-30.95}\\
            \hline
            100 Hz & -27.71 & -30.06 & -29.15 & -30.80 & \textbf{-31.04}\\
            \hline
            130 Hz & -28.89 & -26.96 & -29.23 & -30.82 & \textbf{-31.22}\\
            \hline
            150 Hz & -29.64 & -29.91 & -29.30 & -30.65 & \textbf{-31.04}\\
            \hline
            210 Hz & \textbf{-31.76} & -30.76 & -29.19 & -30.62 & -30.80\\
            \hline
            270 Hz & \textbf{-30.66} & -28.61 & -29.12 & -30.33 & -30.44\\
            \hline
            300 Hz & -29.68 & -30.18 & -29.27 & -30.20 & \textbf{-30.22}\\
            \hline
            400 Hz & \textbf{-30.24} & -29.98 & -28.15 & -29.48 & -29.38\\
            \hline
            500 Hz & \textbf{-29.55} & -28.85 & -27.90 & -28.72 & -28.63\\
            \hline
            800 Hz & \textbf{-26.70} & -18.75 & -25.59 & -26.47 & -26.55\\
            \hline
            1300 Hz & -21.65 & -17.59 & -22.01 & -22.85 & \textbf{-23.24}\\
            \hline
            1850 Hz & -16.86 & -15.25 & -18.29 & -19.18 & \textbf{-19.67}\\
            \hline
        \end{tabular} \\[0.5ex]
        \caption{\textnormal{Performance of \ac{HKF} \& \ac{HKF}-2 (see \ref{sub_sec:model_capacity}) on a range of Doppler values compared against classical and standalone \ac{NN} based baselines.}}
        \label{tab:eval_table}
        \vspace{-5mm}
    \end{table}
    
    \subsection{Impact of increasing model capacity}
    \label{sub_sec:model_capacity}
    In this experiment, we study the effects of increasing the depth (complexity) of the \ac{HKF}. We define \ac{HKF}-2 as \ac{HKF} but with $2$ layers of LSTM. In Table.~\ref{tab:eval_table}, we observe that increasing the depth of \ac{HKF} leads to consistent performance boost for $13$ out of $15$ Doppler frequencies. The result suggests that further improvements in the performance can be achieved by increasing the depth of the network.
    
    \subsection{Doppler interpolation}
        In the following experiment, we evaluate the ability of different tracking schemes to interpolate to unseen Doppler scenarios, i.e., Doppler frequencies that are not present in the training dataset.
        For this experiment, we generate data for one extra Doppler per bin. In case of \ac{BKF}, we use the genie Doppler information to pick the right bin for each test instance, i.e., we choose the \ac{KF} belonging to the exact bin to which the test channel belongs. Depicted in Table.~\ref{tab:eval_table_unseen_dopplers}, our \ac{HKF} and \ac{HKF}-2 outperform LSTM on all the $5$ Doppler frequencies and outperform \ac{BKF} at $4$ out of $5$ cases without genie information. 
        \begin{table}[t]
        \vspace{5mm}
            \centering
                \begin{tabular}{||c|c|c|c|c||} 
                \hline
                \multicolumn{5}{|c|}{MNSE (in dB)} \\
                \hline
                Doppler & BKF & LSTM & HKF-1 & HKF-2 \\ [0.5ex] 
                \hline\hline
                50 Hz & -30.06 & -25.58 & -30.89 & \textbf{-31.12}\\ 
                \hline
                120 Hz & -28.12 & -27.99 & -30.62 & \textbf{-31.05}\\
                \hline
                240 Hz & -29.87 & -28.90 & -30.49 & \textbf{-30.64}\\
                \hline
                450 Hz & \textbf{-29.65} & -27.63 & -29.18 & -29.07\\
                \hline
                1500 Hz & -17.99 & -19.52 & -20.94 & \textbf{-21.27}\\
                \hline
            \end{tabular} \\[0.5ex]
            \caption{\textnormal{Evaluation on untrained Doppler values}}
            \label{tab:eval_table_unseen_dopplers}
        \vspace{-5mm}
        \end{table}
        
        This result is in agreement with our postulate that the \ac{HKF} brings together the flexibility of the RNNs with out of domain generalization capability of the \ac{KF}. The proposed \ac{HKF} demonstrates better Doppler interpolation properties than both the classical baseline and solely \ac{NN} based baseline. In the current experiment, the \ac{HKF}-2 consistently outperforms the \ac{HKF}-1, hence, we use \ac{HKF}-2 for the next experiments. 
        
    \subsection{Generalization properties}
        In the next experiments, we investigate the performance of different schemes when extrapolated to settings beyond the training scenario. For the sake of completeness, we introduce \ac{HKF}-G, a global version of \ac{HKF}-2. The \ac{HKF}-G shares the same model with \ac{HKF}-2, but unlike \ac{HKF}-2 which is trained on a single SNR value of $10$ dB and a pilot ratio of $1:6$, \ac{HKF}-G is trained 
        on SNR values uniformly sampled from the range $5-15$ dB and pilot ratios uniformly sampled from the set $\{1:3, 1:5, 1:6, 1:8, 1:10\}$. The idea is to see how much gain we get by training on a range of different scenarios.

    \subsubsection{Generalization to untrained pilot ratios}
    We start with generalization to different untrained pilot ratios.
    Table.~\ref{tab:eval_table_untrained_pilot_3} shows the evaluation results when both the LSTM and \ac{HKF}-2, trained on a pilot ratio of $1:6$, are evaluated on a pilot ratio of $1:3$. An increased pilot frequency means more frequent observations of the underlying channel. Intuitively, increase in pilot frequency should lead to better channel estimates. As depicted in upper half of  Table.~\ref{tab:eval_table_untrained_pilot_3}, the performance of \ac{HKF}-2 improves by increasing the pilot frequency and it still remains competitive to the \ac{BKF}. On the other hand, despite the more frequent observations, the LSTM completely collapses. 
    
    To increase the scope of our findings, we also evaluated our methods on unseen Doppler values combined with an untrained pilot ratio. The lower half of  Table.~\ref{tab:eval_table_untrained_pilot_3} depicts the results when different schemes are evaluated on untrained Doppler values combined with an untrained pilot ratio of $1:3$. For both the seen and unseen Doppler scenarios, the \ac{HKF}-2 performs competitive to the \ac{BKF} while the LSTM fails to generalize. The \ac{HKF}-G, which includes this pilot ratio in the training phase, performs the best among others.
    \begin{table}[ht]
    \vspace{5mm}
        \centering
            \begin{tabular}{||c|c|c|c|c||} 
            \hline
            \multicolumn{5}{|c|}{MNSE (in dB)}\\
            \hline
            \multicolumn{5}{|c|}{Seen Doppler values}\\
            \hline
            Doppler & BKF & LSTM & HKF-2 & HKF-G \\ [0.5ex] 
            \hline\hline
            0 Hz & \textbf{-35.33} & -18.63 & -34.77 & -35.15\\ 
            \hline
            30 Hz & \textbf{-34.80} & -20.64 & -33.42 & -33.91\\
            \hline
            60 Hz & -31.33 & -20.21 & -33.11 & \textbf{-33.47}\\
            \hline
            70 Hz & -31.96 & -20.16 & -33.11 & \textbf{-33.39}\\
            \hline
            100 Hz & -33.04 & -18.67 & -33.60 & \textbf{-33.93}\\
            \hline
            130 Hz & -30.48 & -17.27 & -33.53 & \textbf{-33.74}\\
            \hline
            150 Hz & -33.09 & -16.16 & -33.41 & \textbf{-33.62}\\
            \hline
            210 Hz & -33.40 & -14.50 & -33.39 & \textbf{-33.63}\\
            \hline
            270 Hz & -32.19 & -13.18 & -32.94 & \textbf{-33.15}\\
            \hline
            300 Hz & \textbf{-33.19} & -12.65 & -32.72 & -32.95\\
            \hline
            400 Hz & \textbf{-32.90} & -11.33 & -32.08 & -32.36\\
            \hline
            500 Hz & \textbf{-32.26} & -10.57 & -31.51 & -31.90\\
            \hline
            800 Hz & -28.06 & -10.52 & -29.61 & \textbf{-30.43}\\
            \hline
            1300 Hz & -27.18 & -9.83 & -27.07 & \textbf{-28.45}\\
            \hline
            1850 Hz & -25.30 & -7.60 & -24.17 & \textbf{-26.53}\\
            \hline
            \multicolumn{5}{|c|}{Unseen Doppler values}\\[0.5ex]
            \hline
            50 Hz & -32.60 & -20.66 & -33.27 & \textbf{-33.68}\\ 
            \hline
            120 Hz & -31.51 & -17.68 & -33.63 & \textbf{-33.89}\\
            \hline
            240 Hz & -32.87 & -13.70 & -33.18 & \textbf{-33.40}\\
            \hline
            450 Hz & \textbf{-32.69} & -10.90 & -31.85 & -32.15\\
            \hline
            1500 Hz & \textbf{-27.28} & -7.71 & -25.40 & -27.04\\
            \hline
        \end{tabular} \\[0.5ex]
        \caption{\textnormal{Evaluation on untrained pilot ratio of $1:3$}}
        \label{tab:eval_table_untrained_pilot_3}
        \vspace{-5mm}
    \end{table}

    
    In the preceding experiment, we experimented with an increased pilot frequency of $1:3$. In the following experiment, we perform a similar experiment but with a reduced pilot frequency of $1:10$. Intuitively, a decrease in pilot frequency would lead to degradation in the performance as there are less frequent observations available to estimate the channel. Table.\ref{tab:eval_table_untrained_pilot_10} depicts the results when different schemes are evaluated on an untrained pilot ratio of $1:10$ combined with seen and unseen Doppler instances. Similar to the previous results, the \ac{HKF}-2 remains competitive to the \ac{BKF} while the LSTM collapses. Owing to its more diverse training, the \ac{HKF}-G fares much better than the other tracking methods. 
    \begin{table}[ht]
    \vspace{5mm}
        \centering
        \begin{tabular}{||c|c|c|c|c||} 
            \hline
            \multicolumn{5}{|c|}{MNSE (in dB)} \\
            \hline
            \multicolumn{5}{|c|}{Seen Doppler values}\\
            \hline
            Doppler & BKF & LSTM & HKF-2 & HKF-G\\ [0.5ex] 
            \hline\hline
            0 Hz & \textbf{-28.56} & -4.55 & -20.44 & -26.16\\ 
            \hline
            30 Hz & \textbf{-31.70} & -5.55 & -19.52 & -26.93\\
            \hline
            60 Hz & -25.80 & -4.85 & -18.62 & \textbf{-27.29}\\
            \hline
            70 Hz & \textbf{-26.16} & -4.31 & -20.68 & -25.30\\
            \hline
            100 Hz & \textbf{-27.59} & -3.29 & -23.47 & -27.39\\
            \hline
            130 Hz & -23.34 & -2.11 & -23.56 & \textbf{-27.17}\\
            \hline
            150 Hz & -26.00 & -1.34 & -25.07 & \textbf{-27.63}\\
            \hline
            210 Hz & \textbf{-27.90} & -0.32 & -24.87 & -27.60\\
            \hline
            270 Hz & -24.41 & 0.34 & -25.54 & \textbf{-27.22}\\
            \hline
            300 Hz & -26.77 & 0.54 & -21.16 & \textbf{-27.09}\\
            \hline
            400 Hz & \textbf{-26.86} & 1.07 & -24.04 & -25.95\\
            \hline
            500 Hz & \textbf{-24.77} & 1.34 & -23.30 & -24.71\\
            \hline
            800 Hz & -10.36 & 1.95 & -18.95 & \textbf{-20.80}\\
            \hline
            1300 Hz & -9.50 & 2.19 & -10.90 & \textbf{-15.45}\\
            \hline
            1850 Hz & -7.25 & 2.23 & -5.18 & \textbf{-11.10}\\
            \hline
            \multicolumn{5}{|c|}{Unseen Doppler values}\\[0.5ex]
            \hline
            50 Hz & \textbf{-27.74} & -5.11 & -23.34 & -24.83\\ 
            \hline
            120 Hz & -25.32 & -2.39 & -23.31 & \textbf{-27.41}\\
            \hline
            240 Hz & -26.43 & 0.07 & -24.40 & \textbf{-27.38}\\
            \hline
            450 Hz & \textbf{-26.28} & 1.19 & -24.14 & -25.38\\
            \hline
            1500 Hz & -10.21 & 2.22 & -8.08 & \textbf{-12.28}\\
            \hline
        \end{tabular} \\[0.5ex]
        \caption{\textnormal{Evaluation on untrained pilot ratio of $1:10$}}
        \label{tab:eval_table_untrained_pilot_10}
        \vspace{-4mm}
    \end{table}

    
    \subsubsection{Generalization to untrained SNR values}
    In the following experiments, we study the generalization properties of different schemes to different noise levels. Both the LSTM, and the \ac{HKF}s, trained on an SNR of $10$ dB, are evaluated on two different SNRs of $5$ dB and $15$ dB. Table.~\ref{tab:eval_table_untrained_snr_5} shows the finding when different schemes are evaluated on an untrained SNR of $5$ dB with seen and unseen Doppler instances. Evident from the Table.~\ref{tab:eval_table_untrained_snr_5}, the performance of LSTM degrades significantly. On the other hand, the \ac{HKF}-2 performs comparable to the \ac{HKF}-G and even outperforms the \ac{BKF} at certain instances.
    
    \begin{table}[ht]
        \centering
        \begin{tabular}{||c|c|c|c|c||} 
            \hline
            \multicolumn{5}{|c|}{MNSE (in dB)} \\
            \hline
            \multicolumn{5}{|c|}{Seen Doppler values}\\
            \hline
            Doppler & BKF & LSTM & HKF-2  & HKF-G\\ [0.5ex] 
            \hline\hline
            0 Hz & -28.62 & -15.83 & -27.15 & \textbf{-28.84}\\ 
            \hline
            30 Hz & \textbf{-29.81} & -17.18 & -27.30 & -28.44\\
            \hline
            60 Hz & -25.08 & -18.46 & -27.79 & \textbf{-28.65}\\
            \hline
            70 Hz & -25.67 & -18.11 & -27.85 & \textbf{-28.56}\\
            \hline
            100 Hz & -26.56 & -18.81 & -27.92 & \textbf{-28.57}\\
            \hline
            130 Hz & -23.50 & -18.64 & -27.99 & \textbf{-28.47}\\
            \hline
            150 Hz & -26.88 & -18.83 & -27.81 & \textbf{-28.31}\\
            \hline
            210 Hz & -27.32 & -19.26 & -27.54 & \textbf{-28.13}\\
            \hline
            270 Hz & -25.59 & -19.81 & -27.07 & \textbf{-27.69}\\
            \hline
            300 Hz & -26.83 & -19.68 & -26.79 & \textbf{-27.50}\\
            \hline
            400 Hz & -26.57 & -19.44 & -25.92 & \textbf{-26.78}\\
            \hline
            500 Hz & -25.63 & -19.52 & -25.13 & \textbf{-26.16}\\
            \hline
            800 Hz & -17.90 & -19.42 & -23.64 & \textbf{-24.19}\\
            \hline
            1300 Hz & -16.77 & -16.94 & -20.93 & \textbf{-21.31}\\
            \hline
            1850 Hz & -14.62 & -14.83 & -18.20 & \textbf{-18.31}\\
            \hline
            \multicolumn{5}{|c|}{Unseen Doppler values}\\[0.5ex]
            \hline
            50 Hz & -26.61 & -18.17 & -27.74 & \textbf{-28.64}\\ 
            \hline
            120 Hz & -24.64 & -18.59 & -27.91 & \textbf{-28.49}\\
            \hline
            240 Hz & -26.58 & -19.18 & -27.29 & \textbf{-27.79}\\
            \hline
            450 Hz & -26.28 & -19.58 & -25.55 & \textbf{-26.40}\\
            \hline
            1500 Hz & -17.17 & -14.85 & -19.23 & \textbf{-19.37}\\
            \hline
        \end{tabular} \\[0.5ex]
        \caption{\textnormal{Evaluation on untrained SNR of $5$ dB}}
        \label{tab:eval_table_untrained_snr_5}
            \vspace{-4mm}
    \end{table}
    
    Next, we evaluate different methods on an untrained SNR of $15$ dB. Tables.~\ref{tab:eval_table_untrained_snr_15} charts the findings of our experiments when different schemes are evaluated on seen and unseen Doppler scenarios. Similar to the prior experiment, the LSTM fails to generalize to a new SNR while the \ac{HKF}-G performs the best. The \ac{HKF}-2, despite being trained on a single SNR ($10$ dB) far away from the test SNRs ($5$ dB and $10$ dB), performs competitive to the \ac{HKF}-G and \ac{BKF}. The \ac{HKF}-2 performs comparable to \ac{BKF} in low-to-mid frequency range and outperforms it at higher frequency range.

    \begin{table}[ht]
        \vspace{5mm}
        \centering
            \begin{tabular}{||c|c|c|c|c||} 
            \hline
            \multicolumn{5}{|c|}{MNSE (in dB)} \\
            \hline
            \multicolumn{5}{|c|}{Seen Doppler values}\\
            \hline
            Doppler & BKF & LSTM & HKF-2 & HKF-G\\ [0.5ex] 
            \hline\hline
            0 Hz & \textbf{-34.00} & -18.76 & -33.70 & -33.26\\ 
            \hline
            30 Hz & \textbf{-34.56} & -22.61 & -31.84 & -32.01\\
            \hline
            60 Hz & -31.08 & -27.59 & -32.11 & \textbf{-32.34}\\
            \hline
            70 Hz & -30.96 & -29.33 & -32.24 & \textbf{-32.44}\\
            \hline
            100 Hz & \textbf{-32.84} & -31.31 & -32.16 & -32.63\\
            \hline
            130 Hz & -29.53 & -31.72 & -32.59 & \textbf{-32.89}\\
            \hline
            150 Hz & -32.29 & -30.41 & -32.46 & \textbf{-32.65}\\
            \hline
            210 Hz & \textbf{-33.98} & -30.74 & -32.24 & -32.62\\
            \hline
            270 Hz & -31.05 & -29.89 & -31.82 & \textbf{-32.08}\\
            \hline
            300 Hz & \textbf{-33.28} & -31.56 & -31.55 & -31.90\\
            \hline
            400 Hz & \textbf{-33.22} & -29.48 & -30.59 & -31.01\\
            \hline
            500 Hz & \textbf{-31.71} & -29.95 & -29.69 & -30.12\\
            \hline
            800 Hz & -19.11 & -23.72 & \textbf{-27.74} & -27.72\\
            \hline
            1300 Hz & -17.95 & -23.10 & \textbf{-24.18} & -23.92\\
            \hline
            1850 Hz & -15.52 & -18.98 & \textbf{-20.08} & -19.62\\
            \hline
            \multicolumn{5}{|c|}{Unseen Doppler values}\\[0.5ex]
            \hline
            50 Hz & \textbf{-32.78} & -27.08 & -32.44 & -32.59\\ 
            \hline
            120 Hz & -30.79 & -29.08 & -32.28 & \textbf{-32.74}\\
            \hline
            240 Hz & \textbf{-32.79} & -30.36 & -32.05 & -32.37\\
            \hline
            450 Hz & \textbf{-32.80} & -29.14 & -30.20 & -30.62\\
            \hline
            1500 Hz & -18.35 & -20.15 & \textbf{-21.82} & -21.29\\
            \hline
        \end{tabular} \\[0.5ex]
        \caption{\textnormal{Evaluation on untrained SNR of $15$ dB}}
        \label{tab:eval_table_untrained_snr_15}
            \vspace{-5mm}
    \end{table}

\section{Conclusion}
Exclusive neural network based solutions to channel tracking problem generalize poorly to unseen scenarios with significant performance degradation. On the other hand, although Kalman filter generalizes well to different SNR and observation patterns, it needs to be adapted continually to changing channel dynamics. We propose a hypernetwork Kalman filter solution which reconciles the best part of both approaches. It was shown that a single \ac{HKF} could be used on a wide range of Doppler values with good out of domain generalization. We believe these neural augmentation approaches are well suited for maximal utilization of the domain knowledge.
\section*{Acknowledgment}
The authors would like to thank Pouriya Sadeghi and Supratik Bhattacharjee for many fruitful discussions. 

\bibliographystyle{IEEEtran}
\bibliography{references.bib}
\end{document}